\documentclass[11pt]{article}

\usepackage{acl} 

\usepackage{times}
\usepackage{latexsym}

\usepackage[T1]{fontenc}

\usepackage[utf8]{inputenc}

\usepackage{microtype}

\usepackage{inconsolata}

\usepackage{graphicx}

\title{Beyond Monoliths: Expert Orchestration for More Capable, Democratic, and Safe Language Models}

\author{
\textbf{Philip Quirke\textsuperscript{1}\thanks{Corresponding author: \texttt{\url{philip@withmartian.com}}}},
\textbf{Narmeen Oozeer\textsuperscript{1}},
\textbf{Chaithanya Bandi\textsuperscript{1}},
\textbf{Amir Abdullah\textsuperscript{1,2}},\\
\textbf{Jason Hoelscher-Obermaier\textsuperscript{3}},
\textbf{Jeff M. Phillips\textsuperscript{4}},
\textbf{Joshua Greaves\textsuperscript{1}},
\textbf{Clement Neo\textsuperscript{5}},\\
\textbf{Michael Lan\textsuperscript{1}},
\textbf{Fazl Barez\textsuperscript{1,6,7}},
\textbf{Shriyash Upadhyay\textsuperscript{1}} \\
\\
\textsuperscript{1}Martian 
\textsuperscript{2}Thoughtworks 
\textsuperscript{3}Apart Research 
\textsuperscript{4}University of Utah \\
\textsuperscript{5}Nanyang Technological University
\textsuperscript{6}University of Oxford 
\textsuperscript{7}WhiteBox
}

\usepackage{wrapfig}  

\usepackage{hyperref}
\usepackage{comment}

\newcommand{\para}[1]{\noindent\textbf{#1.} }

\begin{document}
\maketitle
\vspace{0.5cm}
\begin{abstract}
This position paper argues that the prevailing trajectory toward ever larger, more expensive generalist foundation models controlled by a handful of companies limits innovation and constrains progress. 
We challenge this approach by advocating for an ``Expert Orchestration" (EO) framework as a superior alternative that democratizes LLM advancement. 
Our proposed framework intelligently selects from many existing models based on query requirements and decomposition, focusing on identifying what models do well rather than how they work internally. 
Independent ``judge" models assess various models' capabilities across dimensions that matter to users, while ``router" systems direct queries to the most appropriate specialists within an approved set. 
This approach delivers superior performance by leveraging targeted expertise rather than forcing costly generalist models to address all user requirements. 
EO enhances transparency, control, alignment, performance, safety and democratic participation through intelligent model selection.
\end{abstract}

\section{Introduction}

\begin{figure}[t]
  \includegraphics[width=\columnwidth]{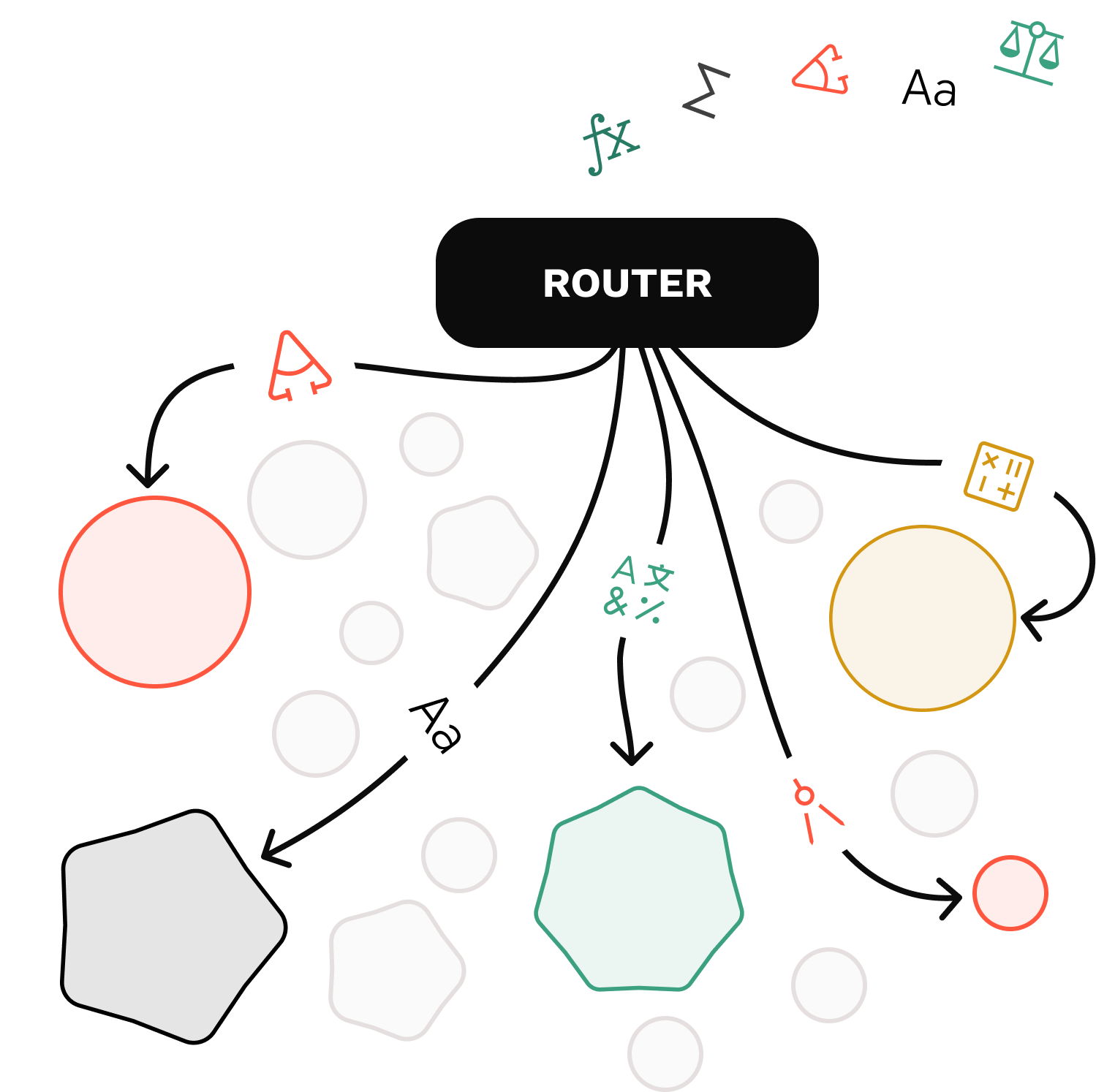}
  \caption{Expert Orchestration framework: The router leverages cached model capability analyses to categorize incoming prompts and dynamically route queries to the optimal model — whether Transformer-based or alternative architectures, large or small, specialist or generalist, cloud-based or edge-deployed — maximizing response quality while minimizing cost.}
  \label{fig:ExpertOrchestration}
\end{figure}

The field of artificial intelligence (AI) has witnessed remarkable progress, largely driven by advancements in large language models (LLMs). Currently, users predominantly rely on monolithic frontier LLMs for queries and tasks. When these models fall short by producing hallucinations \citep{simhi2025trust, zhang2023sirens}, showing bias \citep{gallegos2024bias}, or lacking specialized knowledge \citep{kandpal2022large} the typical response from both developers and users has been to attempt to ``fix" these shortcomings through techniques like prompt engineering, RLHF, vector steering, or parameter-efficient fine-tuning. 
Unfortunately, task interference may cause ``whack-a-mole" side effects compromising a large scale generalist model, as shown by researchers in studies for fine-tuning \citep{shumailov2023curse}, steering \citep{steering_side_effects}, RLHF \citep{rlhf_damage}, and poor prompt selection  \citep{worst_prompt}.

We believe that it is fundamentally intractable ~\citep{varoquaux2025hype} to develop a single model capable of optimal performance across all possible tasks. Hence, this paper argues against the prevailing approach of building ever-larger generalist models. Patching individual models to perform well across all domains is akin to forcing a general practitioner to perform brain surgery rather than deferring the task to a neurosurgeon. Just as humans instinctively consult different experts based on their specific competencies, we should focus on identifying and leveraging the strengths of specialized models. Such specialization not only mirrors natural expertise distribution in human communities, but also offers a more effective path to addressing the limitations of current AI systems.

There are thousands of specialized models currently available on platforms like HuggingFace \citep{huggingface_atlas}. By evaluating their strengths, we can leverage the diversity and specialization of these models. While both generalist and specialist models may require improvements, fixing specialist models is fundamentally more tractable for several reasons. First, specialist models operate in constrained domains with clearer evaluation metrics and more easily established ground truth. Second, the narrower input space dramatically reduces the testing matrix needed to ensure quality. Third, specialized human expertise can be more effectively applied to the limited domain. Finally, when specialists are improved, the benefits immediately propagate through the orchestration system without disrupting other domains, unlike monolithic models where fixes for one domain often cause regressions in others due to parameter interference \citep{forgetting_1, forgetting_2, forgetting_3}. At the same time, \citet{compound-ai-blog} argues state-of-the-art AI performance is increasingly driven not by scaling individual models, but by assembling compound systems composed of multiple coordinated components.

{\bf We posit that improvements in monolithic models aiming to handle all tasks is unsustainable.  
Rather, we propose a paradigm shift towards a framework we term \emph{expert orchestration} (EO), comprised of specialized components: \emph{Judges} that evaluate model capabilities across dimensions that matter to users (factuality, domain expertise, ethics, creativity, etc.), and \emph{Routers} directing user queries to the most appropriate model(s) in a set of specialist \& generalist models, based on user preferences. 
This approach improves control and monitoring, delivering superior answers at lower average cost creating a capable, democratic, and safe ecosystem.   }
 
Below, we outline limitations of the current landscape (Section 2),
why interdisciplinary frameworks argue for change (Section 3),
describe the EO framework (Section 4),
argue it enhances LLM utility (Section 5),
while acknowledging open research questions (Section 6)
and alternative viewpoints exist (Section 7).
Finally we urge adoption to secure a safer AI future (Section 8).

\section{The Problem with Concentrating AI}
\citet{master_switch} describes the recurring cycle, where information industries— such as radio and the internet— begin in a period of innovation but become consolidated by monopolies, which may suppress competition and innovation \citep{stifle_competition}. We highlight the growing risks of similar AI concentration, and call for scrutiny of democratic alternatives by the research community.

\subsection{Market Dynamics}
The economics of frontier AI development are increasingly shaped by powerful market incentives that favor ``winner-take-all" outcomes, where a small number of dominant players capture the lion's share of profits and influence. These companies operate with the expectation that the developers of the most capable generalist LLMs will capture the vast majority of the market, pursuing this consolidation with the potential to concentrate trillions of dollars and substantial deal-making power within a limited number of corporations.



This concentration raises concerns about equity and democratic governance, with potential systemic risks from such market dominance. International governments increasingly view AI concentration in a handful of US companies as a threat to sovereignty, local culture, and democratic values \citep{lecun2024lex}—a concern that extends beyond any single nation's interests to the global distribution of AI capabilities.

EO addresses aspects of these market failures by lowering the resource threshold for meaningful contribution to the AI ecosystem and by creating a framework that naturally incorporates and highlights specialized excellence, regardless of the model creator's scale or resources.

Current market dynamics create concerning misalignments regarding safety. Frontier companies, while often expressing commitment to safety, face real incentives to under-evaluate and under-report potential risks to expedite model releases. This rush to market is driven by intense competition and the desire to capture market share in the perceived ``winner-take-all" dynamic. The DarkBench paper \citep{kran2025darkbench} demonstrates that models misrepresent their own capabilities and advantages over competitors, further complicating accurate risk assessment.

The underlying competitive dynamics often favor rapid capability advancement and market share capture over meticulous safety assurance \citep{martian2025safety}. For companies selling access to increasingly powerful models, the motivation may be weak to dedicate significant resources to in-depth safety research that could slow capability advancements.

This concentration raises concerns about equity and democratic governance, with potential systemic risks from such market dominance.

EO addresses aspects of these market failures by lowering the resource threshold for meaningful contribution to the AI ecosystem and by creating a framework that naturally incorporates and highlights specialized excellence, regardless of the model creator's scale or resources.

\subsection{Technical Challenges of monoliths}

\para{Limited User Insight into LLM ``Thinking" Characteristics}
Beyond the technical correctness, users are increasingly concerned with a range of underlying ``thinking" characteristics. These include legality, morality, the absence of hallucinations, and the lack of gender or other biases. Currently, users have limited means to effectively communicate these criteria to LLMs and possess very limited ability to evaluate how well these models align with their desired thinking characteristics.

Frontier LLMs, however, often present themselves as being universally capable, without any clear differentiation regarding underlying thinking characteristics. While users can gain some limited control over these characteristics by their choice of LLM, and by employing specialized prompts, the actual impact and reliability of these methods remain uncertain.

This limitation is widely recognized in the alignment literature, where recent work emphasizes the importance of user-steerable LLMs and controllable generation. For instance, \citet{bai2022training} introduce Helpful and Harmless Assistant (HH-RLHF), where preferences are directly integrated into model behavior via human feedback loops and fine-tuning procedures.

Similarly, OpenAI's InstructGPT paper \citep{ouyang2022training} shows that aligning LLMs with user intent through instruction-following dramatically improves user satisfaction and safety. However, these efforts are largely global alignment efforts so users do not have fine-grained, per-query control.

Users deserve more direct insight and specific control over the ``thinking" characteristics of LLM behavior. None of the above methods match the explicit and modular control enabled by EO where each thinking characteristic (e.g., legality, bias, hallucinations) is explicitly evaluated and can be chosen by the user per query.

\para{Monolithic Systems Are Less Controllable}
While specialized models and frameworks that enable calling multiple models as tools do exist, ease of use considerations often lead most users to opt for a single LLM, with its inherent strengths and weaknesses, for all their queries.

This single LLM presents as a ``monolith" that is sufficiently proficient across all query types. While this might hold true on average, it is demonstrably false at the individual query level. For many queries, other LLMs, potentially with specialist abilities directly relevant to the query, would be more suitable. Alternatively, a query might be simple enough (e.g., a basic arithmetic problem) that invoking a frontier model represents a wasteful expenditure of resources: money, time, and electricity.

The shift from monoliths to components also mirrors the move in NLP and CV towards modular sparse systems \citep{riquelme2021scaling} and BASE Layers \citep{lewis2021base}, which show that task-specific experts outperform generalist models at lower cost and complexity.

The Modular Deep Learning paper \citep{pfeiffer2023modular} says ``It remains unclear how to develop models that specialize towards multiple tasks without incurring negative interference and that generalize systematically to non-identically distributed tasks". The paper promotes modular deep learning as a potential partial solution to these challenges.

\para{Monolithic Systems Are Cloud-based}
The ever-increasing size of LLMs creates an inherent dependency on cloud infrastructure, ignoring the computational capabilities now present in edge devices where many user queries originate. Cloud-based LLMs rely on a stable network connection for inference, and the response time depends on network stability and speed. When inference occurs locally on edge devices, response time is significantly reduced, and applications can function with limited or no network connectivity \cite{tiwari_edge_2024}. This cloud-centric approach adds unnecessary latency for simple queries, wastes electricity through redundant data transmission, and creates privacy concerns by requiring sensitive user data to travel to remote servers. Google AI Edge has demonstrated that on-device small language models can be effectively deployed on Android, iOS, and Web platforms \cite{google_edge_2025}. These developments challenge the assumption that all AI inference must occur in the cloud.

EO naturally enables a hybrid approach where edge-device routers can evaluate whether user queries need escalation to cloud LLMs or can be handled by specialized SLMs installed locally. The deployment of SLMs on edge devices has emerged as a pivotal strategy to overcome cloud dependency, with organizations ranging from startups to tech giants recognizing this approach---NVIDIA argues that small language models represent the future of agentic AI, while successful deployments on devices like Raspberry Pi and Jetson Nano demonstrate that even resource-constrained hardware can achieve considerable performance improvements without compromising privacy or efficiency \cite{belcak2025smalllanguagemodelsfuture, premai_edge_2025}. 
Users could install specialist SLMs for topics they commonly query, allowing edge-device routers to handle routine requests locally while preserving cloud resources for complex tasks.

\section{Specialization Works: Lessons from NLP and Other Fields}
A growing body of research demonstrates that specialized large language and vision–language models consistently outperform their generalist counterparts when tasks demand domain expertise, structured reasoning, or cultural grounding.
For example, 
domain-trained models such as MatSciBERT \citep{specialization_materials_matscibert}, Me-LLaMA \citep{specialization_medical_mellama}, and LawLLM \citep{specialization_legal_lawllm} surpass larger foundation models on corpora in materials science, medicine, and law.
Even in abstract reasoning domains, WizardMath and mechanistic analyses of arithmetic circuits \citep{specialization_math_wizardmath,specialization_reasoning_arithmetic_quirke} show that compact, fine-tuned models can outperform broader architectures on specialized symbolic tasks.
At the cultural and linguistic level, CultureLLM, LLM-jp, and AfriBERTa \citep{specialization_culture_culturellm,specialization_japanese_llmjp,specialization_african_afriberta} demonstrate that regional and multilingual adaptation improves coherence, fidelity, and social grounding across diverse societies.

Our position in EO builds directly on these findings: specialization is not an anomaly but a general principle observed across complex systems.
Hayek’s theory of distributed knowledge \citep{hayek1945use} recognized that knowledge exists as “dispersed bits” across agents, favoring coordination over centralization.
Smith’s division of labor \citep{smith1776wealth} and Ricardo’s comparative advantage \citep{ricardo1817principles} formalized the same insight—overall efficiency increases when tasks are partitioned according to relative strengths, precisely what EO operationalizes through intelligent routing among models.
Organizational and cognitive theories echo this view: Condorcet’s Jury Theorem \citep{condorcet1785essay} and Hong and Page’s diversity theorem \citep{hong2004groups} show that diverse agents collectively outperform any single expert, while Fodor’s modularity of mind \citep{fodor1983modularity} and Minsky’s society of mind \citep{minsky1986society} describe intelligence as an emergent property of specialized, interacting modules.
Even biological evolution reflects this pattern—the emergence of multicellular organisms with differentiated cell types marks a decisive leap in capability through the division of labor \citep{rueffler2012evolution}.
Orchestration among specialized models similarly enhances collective intelligence: \citet{park2023generative} showed that over a hundred LLM agents, each with distinct roles, collaboratively coordinate a complex social event more efficiently than a single monolithic model.

Taken together, these convergent results across domains affirm EO’s premise: specialization and coordination are enduring principles of intelligent systems, not temporary engineering choices. (See App.~\ref{app:TheoreticalFoundations} for further discussion.)

\section{An Expert Orchestration Framework}
The limitations of monolithic frontier LLMs call for alternative approaches. Here we outline the expert orchestration (EO) framework, a compelling vision designed to overcome several shortcomings.

\para{The Role of Judges}
At the core of EO are specialized models or systems called ``judges" that objectively assess specific characteristics of LLM outputs. For instance, separate judges might evaluate factual accuracy, legal compliance, ethical adherence, or the presence of hallucinations and biases. While judges currently evaluate model responses to user queries, alternative approaches exist — \citet{kadavath2022language} demonstrate that LLMs can assess the validity of their own claims and predict which questions they can answer correctly.

Independent judges enable trust and transparency in EO. By concentrating on distinct evaluation dimensions, judges enable comprehensive assessment of LLM outputs across multiple critical characteristics.

\para{The Role of Routers}
The router receives queries and selects the optimal LLM(s) for each request \citep{prem2025routing}. Routing decisions are informed by judge evaluations and user-specified preferences for response characteristics \citep{towardsdatascience2025routing}. For example, legal questions could be transparently routed to specialized legal models updated with recent case law, rather than relying on generalist models with uncertain legal reasoning (See Fig.\ref{fig:ExpertOrchestration}).

Routers also consider model specialization, operational cost, and response speed \citep{prem2025balancing}. A key advantage is their dynamic nature: EO readily adapts to new LLMs by incorporating them into the model set and using judges to assess their capabilities, enabling continuous evolution. Recent advances in cost-aware routing like HybridLLM and CARROT \citep{ding2024hybridllm, carrot_router}, plus adaptive MoE inference \citep{zhong2024adapmoe} that dynamically selects experts based on task relevance and efficiency tradeoffs, directly support EO implementation.

\begin{figure}[t]
  \includegraphics[width=\columnwidth]{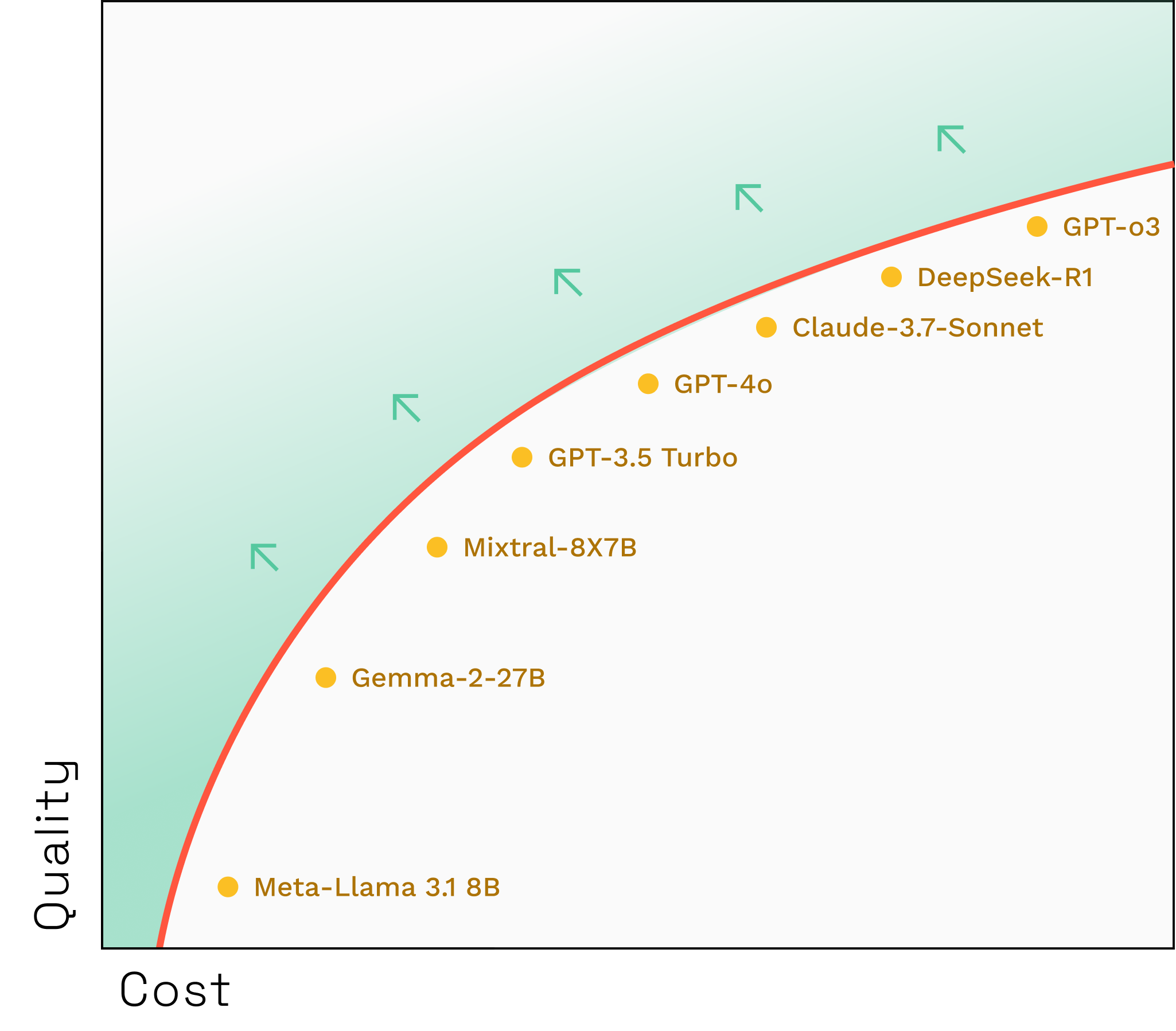}
  \caption{An experimental ``meta-model" (red line) \citep{up-and-to-the-left} uses judges and routers to combine many models. It out-performs any single model (yellow points). More research is expected to move the Quality / Cost pareto curve ``up and to the left" (green arrows). }
  \label{fig:MetaModelPerformanceCurve}
\end{figure}

\para{Superior Performance}
Ensemble methods consistently outperform single models in machine learning \citep{hansen1990neural, dietterich2000ensemble, he2015deep, devlin2018bert} (See Fig.\ref{fig:MetaModelPerformanceCurve}), and this holds for EO through both pre-hoc routing and post-hoc selection approaches.

Pre-hoc routing uses judges to train models that predict which LLM will perform best for each query. CARROT \citep{carrot_router} predicts generation cost and quality from input prompts, achieving higher scores across cost ranges than any single model on RouterBENCH. Similarly, P2L \citep{frick2025p2l}, trained on LMSys ChatBot Arena preference data, topped the leaderboard by predicting user-preferred LLMs per query. Recent work \citep{shnitzer2024universal} even generalizes to unseen models at test time.

Post-hoc methods query multiple LLMs and select the best answer. While approaches like LLMBlender \citep{jiang2023llmblender} are powerful, they incur high costs. More efficient cascading methods like FrugalGPT \citep{chen2023frugalgpt} query models sequentially, returning the first answer exceeding a quality threshold—achieving up to 2.4x cost reduction while maintaining quality. RouterBench experiments \citep{hu2024routerbench} confirm cascades outperform single models across cost levels, with performance heavily dependent on judge quality.

\para{Integration of Specialized Models}
EO enables seamless integration of specialized models into the broader ecosystem \citep{conclusionintelligence2025slms}. When specialized models demonstrate ``best in class" performance in a domain, the router can prioritize them for relevant queries \citep{locaria2025specialized}. This removes the barrier requiring every innovator to match frontier LLM capabilities across all domains to gain market share. Instead, innovators can focus on achieving excellence within a narrower scope \citep{acl2025academics}, fundamentally democratizing model creation and fostering a vibrant community of specialist contributors.

While many specialized models will derive from foundation models, this strengthens our argument. EO realizes the full value of foundation models through selective specialization and strategic routing. Rather than forcing one model to perform optimally across all domains (an impossible task given parameter interference) orchestration leverages foundation capabilities while optimizing performance through specialized deployment. This represents a mature evolution analogous to how early integrated computer systems evolved into specialized components working in concert.

\textbf{Model-Architecture Agnostic. } EO has the advantage of being architecture-agnostic: Beyond transformers, EO can incorporate specialized models like CNNs for vision \citep{zhao_review_2024}, RNNs for streaming data \citep{info15090517}, GNNs for relational structures \citep{ZHOU202057}, and diffusion models for generative tasks \citep{yang2025diffusionmodelscomprehensivesurvey}. For instance, EO may route a graph reasoning task to a GNN, a time-series forecast to an RNN, and a textual explanation to an LLM, then combine outputs through judges. This capacity to integrate diverse architectures increases EO’s utility, as different model architectures may excel in different modalities or constraints such as latency, energy efficiency, or symbolic reasoning. 

\section{How EO Enhances LLM Utility}
EO offers substantial enhancements across several key dimensions of LLM utility, leading to a more robust, user-centric, and responsible ecosystem.

\para{Increased Transparency and Trust} 
A significant benefit of EO lies in its inherent ability to increase transparency and build trust in LLM outputs \citep{ibm2025explainable}. By employing dedicated, independent, and objective judges to evaluate specific characteristics of interest across a multitude of models, the framework provides users with a clearer understanding of the strengths and weaknesses of different LLMs in various domains \citep{smythos2025explainable, pmc2025explainable}. 

EO parallels the rigorous, standardized evaluation practices found in safety-critical domains such as aviation, nuclear energy, and medical device manufacturing, where independent regulatory bodies and engineering frameworks are used to ensure system safety, reliability, and compliance \citep{leveson2016engineering, rushby1994critical,storey1996safety}. Organizations like the Vector Institute and DNV already provide independent evaluations of models and vendors, highlighting the importance of this objective assessment \citep{globenewswire2025vector, dnv2025assessment}. 

Recent work shows the importance of exploring internal reasoning to improve public trust \citep{kook2022deep, thakur2024judging}.

\para{Selection of Judges Empowers Users} 
The use of judges focusing on specific characteristics empowers users with greater control over responses \citep{phenxai2025routing}. 
EO allows a user to specify prioritized characteristics for individual requests, with the router directing queries to models best suited to provide aligned answers \citep{prem2025balancing}.

\para{Decomposing Requests Improves Alignment, Control, and Accuracy}.
EO facilitates decomposing complex requests into manageable steps, such as planning followed by execution phases \citep{eyelevel2025rag}. Specialized project models handle planning, with ``supervisor" models reviewing results to enhance safety. Costing models estimate required resources, while execution steps are delegated to domain-specific models.

This decomposition provides natural monitoring points and reduces the ``scope of control" of any single model, lessening reliance on potentially misaligned models and mitigating single points of failure. Untrustworthy models can be swapped out. Decomposition allows us to start developing robust control techniques now.

\paragraph{Realigning Market Incentives Towards Specialization and Competition.}
EO fundamentally restructures market dynamics by eliminating both the transaction costs and competitive moats that make specialized models economically unviable \citep{varoquaux2025hype}. Currently, users face high switching costs when moving between different models for different tasks—learning new interfaces, managing multiple subscriptions, and remembering which model works best for what. These frictions make generalist models attractive despite inferior performance in specific domains. Simultaneously, incumbent companies build defensive moats by creating models that are ``good enough" across many domains, making it hard for users to justify switching despite superior specialists existing. EO destroys both barriers: it removes transaction costs through seamless automatic routing behind a unified interface, while eliminating defensive moats by automatically choosing the best model for each task. This makes it impossible to defend market position through convenience rather than capability---companies must continuously earn their position through specialized excellence.

Organizations implementing EO face structural incentives that naturally promote ecosystem health. To maximize routing accuracy and customer value, they must maintain comprehensive, objective evaluations of available models on a per domain basis, combating the proliferation of contaminated benchmarks in training datasets \citep{contamination_1, contamination_2} for general capabilities.  In doing so, they are also economically motivated to continuously seek out and integrate the most effective specialized models. This creates sustainable market demand for niche innovators while incentivizing transparency: orchestration providers gain credibility through verifiable model evaluations rather than capability hoarding, creating competitive pressure toward better measurement and disclosure.

Furthermore, the organization is naturally driven to publish objective ``leaderboards" that rank models based on their performance across various capability areas. This transparency provides a clear benchmark for innovators, who then only need to create a model that excels in a specific area to gain recognition and potential integration into an EO implementation. EO stops the possibility of ``best general model captures all value".

\section{Research Directions and Challenges}

EO, while promising, presents several key research questions that represent exciting areas for innovation.  First, developing robust methodologies for evaluating models across diverse ``thinking" characteristics beyond traditional metrics is essential, including bias \citep{llm_bias_testing}, fairness \citep{ai_fairness_metrics}, and hallucination detection \citep{llm_evaluation_metrics}.

Research is needed on utilizing multiple judges that reflect diverse user preferences to inform routing decisions.
Deeply understanding model capabilities beyond simple benchmarking is necessary for optimal task matching.

Additional research directions include: 
(1) developing efficient and scalable routing algorithms that handle numerous models and complex preferences; 
(2) addressing the cold-start problem for new models with limited performance data; 
(3) exploring techniques for composing specialized models \citep{Survery_ModelMerging_2024} to create more powerful capabilities; 
(4) studying broader ecosystem dynamics and impacts on competition and innovation; 
(5) applying dynamic model selection techniques \citep{dynamic_classifier_selection} for adaptive routing; 
(6) developing theoretical models for when and how EO can improve performance such as with boosting; and 
(7) how to leverage ensemble methods \citep{llm_ensemble_survey} and cost-aware routing \citep{carrot_router} to optimize performance and efficiency.

Research into different architectures for implementing judges -- including fine-tuned specialized models, rule-based systems, and human evaluation integration -- represents another critical area for investigation. Together, these research directions will help realize the full potential of EO while addressing its current limitations.

\section{Alternative Views and Considerations}

As with any emerging framework, EO invites thoughtful critique and warrants a balanced evaluation. Several alternative perspectives surfaced during the development of this work:

\textbf{Centralization of Orchestration Infrastructure.} Critics note that EO could shift gatekeeping from models to those controlling routers and judges, potentially recreating winner-take-all dynamics. However, unlike monolithic models, orchestration components are logically separable—enabling regulatory intervention, user customization (bringing own judges or selecting model families), and lower barriers to entry. The heterogeneity of routing needs across domains suggests market fragmentation rather than consolidation. Crucially, transparency requirements are far more enforceable for modular orchestration systems than for opaque monoliths, making democratic oversight tractable.

\textbf{Corporate Incentives and Public Benefit Structures.} This paper emphasizes misaligned incentives in frontier model development. Critics note that some organizations operate as or are transitioning to public benefit corporations (PBCs) and are legally permitted—and in some cases obligated—to prioritize societal welfare alongside shareholder value. This complicates a purely profit-motivated critique.

\textbf{Latency and Cost Trade-offs.} The orchestration of multiple models introduces questions about computational efficiency. Decomposing queries and routing them through specialized evaluators and responders may increase latency or system overhead in certain cases. 
These costs must be weighed against the benefits of specialized performance and may be mitigated through efficient pre-hoc routing strategies.

\textbf{Applicability Beyond Language.} Readers may view EO as specific to LLMs. In practice, the framework generalizes to other modalities—including vision, speech, and multimodal systems—where specialized components can also enhance performance, transparency, and control.

\textbf{Sufficiency of Generalist Models.} Critics may see current generalist models, particularly when augmented with tool use, as “good enough” for most applications. We believe this view underestimates both the current limitations and long-term risks. Specialized systems consistently outperform generalists in high-stakes or knowledge-intensive domains. EO offers structural benefits—transparent governance, distributed safety guarantees, and robust oversight—that generalist architectures and tool use alone cannot provide.

\textbf{Relation to Existing Approaches.}
While EO proposes a shift in paradigm, it is important to situate this framework alongside existing strategies that have shaped the current AI landscape. We consider most of these as orthogonal to EO, and would still have their use cases.

\emph{Scaling Frontier LLMs.} The dominant approach in large labs has been to continually scale generalist models. This delivers strong average performance but at rising economic and environmental costs, with diminishing returns at the margins. EO instead leverages frontier models only where necessary, while enabling smaller specialized models to contribute value, lowering barriers to entry.

\emph{Reinforcement Learning with Human Feedback (RLHF) and Fine-Tuning.} While effective for specialized domains, they often introduce ``whack-a-mole'' regressions in generalist models where improvements in one area degrade others~\cite{ouyang2022training, rlhf_damage}. EO mitigates this brittleness by not requiring a single model to serve all purposes: weaknesses in an expert can be compensated by routing to another without destabilizing the system.

\emph{Agentic AI.} Agentic frameworks expand the action space of a single LLM by enabling planning, tool use, and multi-turn reasoning \citep{singh2025artist}. They increase capability but still rely on one model’s internal judgment for when and how to act. EO instead governs \textit{which} model should act, using independent judges and routers to select the most appropriate specialist. So EO can treat an agentic LLM as just another expert within its ecosystem, subject to routing and oversight.

\emph{Multi-Agentic AI.} Recent work explores ``societies'' of LLM agents that collaborate via debate, role specialization, or simulation \cite{ye2025masgpt,han2025multiagent}. These systems demonstrate emergent capabilities, but typically rely on multiple instantiations of the \emph{same} generalist model and depend on endogenous coordination. As a result, they face persistent challenges in task allocation, layered context and memory management, and ensuring reliable collective decisions \cite{han2025multiagent}. EO reframes this paradigm as a \emph{system-level meta-controller}: it \emph{explicitly} curates heterogeneity by incorporating specialist models, assigns roles through routers, and enforces policy and quality constraints via judges. So EO subsumes some of the benefits of multi-agent interaction while adding verifiable oversight, auditability, and cost/latency controls—providing governance, transparency, and democratic participation.

\emph{Ensembling and Mixture-of-Experts (MoE).} Approaches such as LLMBlender \citep{jiang2023llmblender}, FrugalGPT\citep{chen2023frugalgpt}, and MoE \citep{zhong2024adapmoe} architectures combine multiple outputs or subnetworks to improve performance. EO differs by orchestrating independent models across organizational boundaries, allowing new specialists to enter without retraining the whole system.

These perspectives highlight important avenues for ongoing reflection, implementation caution, and further research, which we believe strengthen rather than diminish the case for EO.

\section{Conclusion: Towards a More Robust and Human-Aligned Future}
The current dominance of monolithic frontier LLMs suffers from inherent limitations related to winner-take-all dynamics, misaligned safety incentives, barriers to entry for specialized models, limited user insight, and the inefficiencies of a one-size-fits-all approach.

EO offers a compelling alternative that addresses these shortcomings by introducing a framework composed of specialized evaluation models (``judges") and intelligent routing systems (``routers"). This approach promises higher quality answers at a lower average cost by strategically leveraging the strengths of diverse models, including both frontier and specialized ones.

The framework enhances transparency and trust through independent evaluation, empowers users with granular control over desired characteristics, improves alignment and accuracy through request decomposition, and fosters a more democratic and open ecosystem. Moreover, an organization implementing EO has incentives naturally aligned with safety and transparency.

If AGI does not emerge suddenly from a single generalist system, an EO framework could achieve AGI earlier than general models. Our approach enables the development now of strong safeguards that decrease potential extinction-level threats.

By addressing many limitations of the current paradigm and offering a path towards a more user-centric and responsible future, EO holds significant promise for shaping the next generation of language model applications.

\bibliography{bibliography}

\begin{thebibliography}{99}
\providecommand{\natexlab}[1]{#1}

\bibitem[{Aghion et~al.(2023)Aghion, Bergeaud, Boppart, Klenow, and Li}]{stifle_competition}
Philippe Aghion, Antonin Bergeaud, Timo Boppart, Peter~J Klenow, and Huiyu Li. 2023.
\newblock A theory of falling growth and rising rents.
\newblock \emph{Review of Economic Studies}, 90(6):2675--2702.

\bibitem[{Aizawa et~al.(2024)Aizawa, Aramaki, Chen, Cheng, Deguchi, Enomoto, Fujii, Fukumoto, Fukushima, Han et~al.}]{specialization_japanese_llmjp}
Akiko Aizawa, Eiji Aramaki, Bowen Chen, Fei Cheng, Hiroyuki Deguchi, Rintaro Enomoto, Kazuki Fujii, Kensuke Fukumoto, Takuya Fukushima, Namgi Han, and 1 others. 2024.
\newblock Llm-jp: A cross-organizational project for the research and development of fully open japanese llms.
\newblock \emph{arXiv preprint arXiv:2407.03963}.

\bibitem[{Bai et~al.(2022)Bai, Jones, Ndousse, Askell, Chen, DasSarma, Drain, Fort, Ganguli, Henighan et~al.}]{bai2022training}
Yuntao Bai, Andy Jones, Kamal Ndousse, Amanda Askell, Anna Chen, Nova DasSarma, Dawn Drain, Stanislav Fort, Deep Ganguli, Tom Henighan, and 1 others. 2022.
\newblock Training a helpful and harmless assistant with reinforcement learning from human feedback.
\newblock \emph{arXiv preprint arXiv:2204.05862}.

\bibitem[{Belcak et~al.(2025)Belcak, Heinrich, Diao, Fu, Dong, Muralidharan, Lin, and Molchanov}]{belcak2025smalllanguagemodelsfuture}
Peter Belcak, Greg Heinrich, Shizhe Diao, Yonggan Fu, Xin Dong, Saurav Muralidharan, Yingyan~Celine Lin, and Pavlo Molchanov. 2025.
\newblock \href {https://arxiv.org/abs/2506.02153} {Small language models are the future of agentic ai}.
\newblock \emph{Preprint}, arXiv:2506.02153.

\bibitem[{Brownlee(2025)}]{dynamic_classifier_selection}
Jason Brownlee. 2025.
\newblock \href {https://machinelearningmastery.com/dynamic-classifier-selection-in-python/} {Dynamic classifier selection ensembles in python}.
\newblock \emph{Machine Learning Mastery}.

\bibitem[{Cao et~al.(2024)Cao, Cai, Zhang, Zou, and Lam}]{worst_prompt}
Bowen Cao, Deng Cai, Zhisong Zhang, Yuexian Zou, and Wai Lam. 2024.
\newblock On the worst prompt performance of large language models.
\newblock \emph{arXiv preprint arXiv:2406.10248}.

\bibitem[{Chen et~al.(2023)Chen, Zaharia, and Zou}]{chen2023frugalgpt}
Lingjiao Chen, Matei Zaharia, and James Zou. 2023.
\newblock Frugalgpt: How to use large language models while reducing cost and improving performance.
\newblock \emph{arXiv preprint arXiv:2305.05176}.

\bibitem[{Chen et~al.(2025)Chen, Li, Chen, Li, Sun, Luo, Mao, Yang, Sun, and Yu}]{llm_ensemble_survey}
Zhijun Chen, Jingzheng Li, Pengpeng Chen, Zhuoran Li, Kai Sun, Yuankai Luo, Qianren Mao, Dingqi Yang, Hailong Sun, and Philip~S. Yu. 2025.
\newblock \href {https://arxiv.org/abs/2502.18036} {Harnessing multiple large language models: A survey on llm ensemble}.
\newblock \emph{arXiv preprint arXiv:2502.18036}.

\bibitem[{Christiano et~al.(2018)Christiano, Shlegeris, and Amodei}]{christiano2018supervising}
Paul Christiano, Buck Shlegeris, and Dario Amodei. 2018.
\newblock Supervising strong learners by amplifying weak experts.
\newblock \emph{arXiv preprint arXiv:1810.08575}.

\bibitem[{{Conclusion Intelligence}(2025)}]{conclusionintelligence2025slms}
{Conclusion Intelligence}. 2025.
\newblock The rise of specialized language models (slms).
\newblock \url{https://conclusion.intelligence.com/}.
\newblock Accessed: 2025-05-21.

\bibitem[{Condorcet(1785)}]{condorcet1785essay}
Marquis~de Condorcet. 1785.
\newblock \emph{Essay on the application of analysis to the probability of majority decisions}.
\newblock Paris: De l'imprimerie royale.

\bibitem[{Dahl(2008)}]{dahl2008democracy}
Robert~A Dahl. 2008.
\newblock \emph{Democracy and its Critics}.
\newblock Yale university press.

\bibitem[{Deng et~al.(2023)Deng, Zhao, Tang, Gerstein, and Cohan}]{contamination_2}
Chunyuan Deng, Yilun Zhao, Xiangru Tang, Mark Gerstein, and Arman Cohan. 2023.
\newblock Investigating data contamination in modern benchmarks for large language models.
\newblock \emph{arXiv preprint arXiv:2311.09783}.

\bibitem[{Devlin et~al.(2018)Devlin, Chang, Lee, and Toutanova}]{devlin2018bert}
Jacob Devlin, Ming-Wei Chang, Kenton Lee, and Kristina Toutanova. 2018.
\newblock Bert: Pre-training of deep bidirectional transformers for language understanding.
\newblock \emph{arXiv preprint arXiv:1810.04805}.

\bibitem[{Dewey and Rogers(2012)}]{dewey2012public}
John Dewey and Melvin~L Rogers. 2012.
\newblock \emph{The public and its problems: An essay in political inquiry}.
\newblock Penn State Press.

\bibitem[{Dietterich(2000)}]{dietterich2000ensemble}
Thomas~G Dietterich. 2000.
\newblock Ensemble methods in machine learning.
\newblock \emph{Multiple Classifier Systems}, pages 1--15.

\bibitem[{Ding et~al.(2024)Ding, Mallick, Wang, Sim, Mukherjee, Ruhle, Lakshmanan, and Awadallah}]{ding2024hybridllm}
Dujian Ding, Ankur Mallick, Chi Wang, Robert Sim, Subharata Mukherjee, Victor Ruhle, Laks~VS Lakshmanan, and Ahmed~Hassan Awadallah. 2024.
\newblock Hybrid llm: Cost-efficient and quality-aware query routing.
\newblock \emph{arXiv preprint arXiv:2404.14618}.

\bibitem[{{DNV Group}(2025)}]{dnv2025assessment}
{DNV Group}. 2025.
\newblock Ai vendor capability assessment: Demonstrate trustworthiness of your ai solution.
\newblock \url{https://www.dnv.com/}.
\newblock Accessed: 2025-05-21.

\bibitem[{Dodge et~al.(2021)Dodge, Sap, Marasovi{\'c}, Agnew, Ilharco, Groeneveld, Mitchell, and Gardner}]{contamination_1}
Jesse Dodge, Maarten Sap, Ana Marasovi{\'c}, William Agnew, Gabriel Ilharco, Dirk Groeneveld, Margaret Mitchell, and Matt Gardner. 2021.
\newblock Documenting large webtext corpora: A case study on the colossal clean crawled corpus.
\newblock \emph{arXiv preprint arXiv:2104.08758}.

\bibitem[{Dredze et~al.(2024)Dredze, Winata, Kambadur, Wu, {\.I}rsoy, Lu, Dabravolski, Rosenberg, and Gehrmann}]{acl2025academics}
Mark Dredze, Genta~Indra Winata, Prabhanjan Kambadur, Shijie Wu, Ozan {\.I}rsoy, Steven Lu, Vadim Dabravolski, David Rosenberg, and Sebastian Gehrmann. 2024.
\newblock Academics can contribute to domain-specialized language models.
\newblock In \emph{Proceedings of the 2024 Conference on Empirical Methods in Natural Language Processing}, pages 5100--5110.

\bibitem[{{Eyelevel.ai}(2025)}]{eyelevel2025rag}
{Eyelevel.ai}. 2025.
\newblock Optimizing rag systems with advanced llm routing techniques: A deep dive.
\newblock \url{https://eyelevel.ai/}.
\newblock Accessed: 2025-05-21.

\bibitem[{Fodor(1983)}]{fodor1983modularity}
Jerry~A Fodor. 1983.
\newblock \emph{The modularity of mind: An essay on faculty psychology}.
\newblock MIT press.

\bibitem[{Frick et~al.(2025)Frick, Chen, Tennyson, Li, Chiang, Angelopoulos, and Stoica}]{frick2025p2l}
Evan Frick, Connor Chen, Joseph Tennyson, Tianle Li, Wei-Lin Chiang, Anastasios~N Angelopoulos, and Ion Stoica. 2025.
\newblock Prompt-to-leaderboard.
\newblock \emph{arXiv preprint arXiv:2502.14855}.

\bibitem[{Gallegos et~al.(2024)Gallegos, Rossi, Barrow, Tanjim, Kim, Dernoncourt, Yu, Zhang, and Ahmed}]{gallegos2024bias}
Isabel~O Gallegos, Ryan~A Rossi, Joe Barrow, Md~Mehrab Tanjim, Sungchul Kim, Franck Dernoncourt, Tong Yu, Ruiyi Zhang, and Nesreen~K Ahmed. 2024.
\newblock Bias and fairness in large language models: A survey.
\newblock \emph{Computational Linguistics}, 50(3):1097--1179.

\bibitem[{{GlobeNewswire}(2025)}]{globenewswire2025vector}
{GlobeNewswire}. 2025.
\newblock Vector institute unveils comprehensive evaluation of leading models.
\newblock \url{https://www.globenewswire.com/}.
\newblock Accessed: 2025-05-21.

\bibitem[{{Google AI Edge}(2025)}]{google_edge_2025}
{Google AI Edge}. 2025.
\newblock \href {https://developers.googleblog.com/en/google-ai-edge-small-language-models-multimodality-rag-function-calling/} {On-device small language models with multimodality, {RAG}, and function calling}.
\newblock Google Developers Blog.
\newblock Accessed: 2025-05-20.

\bibitem[{Gupta et~al.(2022)Gupta, Zaki, Krishnan, and Mausam}]{specialization_materials_matscibert}
Tanishq Gupta, Mohd Zaki, NM~Anoop Krishnan, and Mausam. 2022.
\newblock Matscibert: A materials domain language model for text mining and information extraction.
\newblock \emph{npj Computational Materials}, 8(1):102.

\bibitem[{Habermas(2015)}]{habermas2015between}
J{\"u}rgen Habermas. 2015.
\newblock \emph{Between facts and norms: Contributions to a discourse theory of law and democracy}.
\newblock John Wiley \& Sons.

\bibitem[{Han et~al.(2025)Han, Zhang, Yao, Jin, and Xu}]{han2025multiagent}
Shanshan Han, Qifan Zhang, Yuhang Yao, Weizhao Jin, and Zhaozhuo Xu. 2025.
\newblock \href {https://arxiv.org/abs/2402.03578} {Llm multi-agent systems: Challenges and open problems}.
\newblock \emph{arXiv preprint arXiv:2402.03578}.

\bibitem[{Hansen and Salamon(1990)}]{hansen1990neural}
Lars~Kai Hansen and Peter Salamon. 1990.
\newblock Neural network ensembles.
\newblock \emph{IEEE Transactions on Pattern Analysis and Machine Intelligence}, 12(10):993--1001.

\bibitem[{Hayek(1945)}]{hayek1945use}
Friedrich~A Hayek. 1945.
\newblock The use of knowledge in society.
\newblock \emph{The American Economic Review}, 35(4):519--530.

\bibitem[{He et~al.(2015)He, Zhang, Ren, and Sun}]{he2015deep}
Kaiming He, Xiangyu Zhang, Shaoqing Ren, and Jian Sun. 2015.
\newblock Deep residual learning for image recognition.
\newblock \emph{Proceedings of the IEEE Conference on Computer Vision and Pattern Recognition}, pages 770--778.

\bibitem[{Hong and Page(2004)}]{hong2004groups}
Lu~Hong and Scott~E Page. 2004.
\newblock Groups of diverse problem solvers can outperform groups of high-ability problem solvers.
\newblock \emph{Proceedings of the National Academy of Sciences}, 101(46):16385--16389.

\bibitem[{Horwitz et~al.(2025)Horwitz, Kurer, Kahana, Amar, and Hoshen}]{huggingface_atlas}
Eliahu Horwitz, Nitzan Kurer, Jonathan Kahana, Liel Amar, and Yedid Hoshen. 2025.
\newblock Charting and navigating hugging face's model atlas.
\newblock \emph{arXiv preprint arXiv:2503.10633}.

\bibitem[{Hu et~al.(2024)Hu, Bieker, Li, Jiang, Keigwin, Ranganath, Keutzer, and Upadhyay}]{hu2024routerbench}
Qitian~Jason Hu, Jacob Bieker, Xiuyu Li, Nan Jiang, Benjamin Keigwin, Gaurav Ranganath, Kurt Keutzer, and Shriyash~Kaustubh Upadhyay. 2024.
\newblock Routerbench: A benchmark for multi-llm routing system.
\newblock \emph{arXiv preprint arXiv:2403.12031}.

\bibitem[{{IBM}(2025)}]{ibm2025explainable}
{IBM}. 2025.
\newblock What is explainable ai (xai)?
\newblock \url{https://www.ibm.com/}.
\newblock Accessed: 2025-05-21.

\bibitem[{Irving et~al.(2018)Irving, Christiano, and Amodei}]{irving2018ai}
Geoffrey Irving, Paul Christiano, and Dario Amodei. 2018.
\newblock Ai safety via debate.
\newblock In \emph{International Conference on Learning Representations Workshop}.

\bibitem[{Jiang et~al.(2023)Jiang, Ren, and Lin}]{jiang2023llmblender}
Dongfu Jiang, Xiang Ren, and Bill~Yuchen Lin. 2023.
\newblock Llm-blender: Ensembling large language models with pairwise ranking and generative fusion.
\newblock \emph{arXiv preprint arXiv:2306.02561}.

\bibitem[{Kadavath et~al.(2022)Kadavath, Conerly, Askell, Henighan, Drain, Perez, Schiefer, Dodds, DasSarma, Tran-Johnson et~al.}]{kadavath2022language}
Saurav Kadavath, Tom Conerly, Amanda Askell, Tom Henighan, Dawn Drain, Ethan Perez, Nelson Schiefer, Zac Dodds, Nova DasSarma, Eli Tran-Johnson, and 1 others. 2022.
\newblock Language models (mostly) know what they know.
\newblock \emph{arXiv preprint arXiv:2207.05221}.

\bibitem[{Kandpal et~al.(2022)Kandpal, Deng, Roberts, Wallace, and Raffel}]{kandpal2022large}
Nikhil Kandpal, Haikang Deng, Adam Roberts, Eric Wallace, and Colin Raffel. 2022.
\newblock Large language models struggle to learn long-tail knowledge.
\newblock \emph{arXiv preprint arXiv:2211.08411}.

\bibitem[{Kirk et~al.(2023)Kirk, Mediratta, Nalmpantis, Luketina, Hambro, Grefenstette, and Raileanu}]{rlhf_damage}
Robert Kirk, Ishita Mediratta, Christoforos Nalmpantis, Jelena Luketina, Eric Hambro, Edward Grefenstette, and Roberta Raileanu. 2023.
\newblock Understanding the effects of rlhf on llm generalisation and diversity.
\newblock \emph{arXiv preprint arXiv:2310.06452}.

\bibitem[{Kook et~al.(2022)Kook, G{\"o}tschi, Baumann, Hothorn, and Sick}]{kook2022deep}
Lucas Kook, Andrea G{\"o}tschi, Philipp~FM Baumann, Torsten Hothorn, and Beate Sick. 2022.
\newblock Deep interpretable ensembles.
\newblock \emph{arXiv preprint arXiv:2205.12729}.

\bibitem[{Kran et~al.(2025)Kran, Nguyen, Kundu, Jawhar, Park, and Jurewicz}]{kran2025darkbench}
Esben Kran, Jord Nguyen, Akash Kundu, Sami Jawhar, Jinsuk Park, and Mateusz Jurewicz. 2025.
\newblock Darkbench: Benchmarking dark patterns in large language models.
\newblock In \emph{Proceedings of the International Conference on Learning Representations (ICLR)}. ICLR.

\bibitem[{LeCun(2024)}]{lecun2024lex}
Yann LeCun. 2024.
\newblock \href {https://www.lesswrong.com/posts/bce63kvsAMcwxPipX/highlights-from-lex-fridman-s-interview-of-yann-lecun} {Interview with lex fridman}.
\newblock LessWrong.

\bibitem[{Leveson(2016)}]{leveson2016engineering}
Nancy~G Leveson. 2016.
\newblock \emph{Engineering a safer world: Systems thinking applied to safety}.
\newblock The MIT Press.

\bibitem[{Lewis et~al.(2021)Lewis, Bhosale, Dettmers, Goyal, and Zettlemoyer}]{lewis2021base}
Mike Lewis, Shruti Bhosale, Tim Dettmers, Naman Goyal, and Luke Zettlemoyer. 2021.
\newblock Base layers: Simplifying training of large, sparse models.
\newblock \emph{International Conference on Machine Learning}, pages 6265--6274.

\bibitem[{Li et~al.(2024)Li, Chen, Wang, Sitaram, and Xie}]{specialization_culture_culturellm}
Cheng Li, Mengzhuo Chen, Jindong Wang, Sunayana Sitaram, and Xing Xie. 2024.
\newblock Culturellm: Incorporating cultural differences into large language models.
\newblock \emph{Advances in Neural Information Processing Systems}, 37:84799--84838.

\bibitem[{{Locaria}(2025)}]{locaria2025specialized}
{Locaria}. 2025.
\newblock Specialized models rise as llms stall.
\newblock \url{https://locaria.com/}.
\newblock Accessed: 2025-05-21.

\bibitem[{Luo et~al.(2023)Luo, Sun, Xu, Zhao, Lou, Tao, Geng, Lin, Chen, and Zhang}]{specialization_math_wizardmath}
Haipeng Luo, Qingfeng Sun, Can Xu, Pu~Zhao, Jianguang Lou, Chongyang Tao, Xiubo Geng, Qingwei Lin, Shifeng Chen, and Dongmei Zhang. 2023.
\newblock Wizardmath: Empowering mathematical reasoning for large language models via reinforced evol-instruct.
\newblock \emph{arXiv preprint arXiv:2308.09583}.

\bibitem[{Madison(1788)}]{madison1788federalist}
James Madison. 1788.
\newblock The federalist no. 51: The structure of the government must furnish the proper checks and balances between the different departments.
\newblock \emph{Independent Journal}, 6.

\bibitem[{{Martian}(2025{\natexlab{a}})}]{martian2025safety}
{Martian}. 2025{\natexlab{a}}.
\newblock Safety vs capitalism - blog.
\newblock \url{https://blog.withmartian.com/post/ai-safety-incentives/}.
\newblock Accessed: 2025-05-21.

\bibitem[{{Martian}(2025{\natexlab{b}})}]{up-and-to-the-left}
{Martian}. 2025{\natexlab{b}}.
\newblock \href {https://www.withmartian.com/post/up-and-to-the-left} {Up and to the left! how martian uses routing to push the pareto frontier}.
\newblock Martian Blog.
\newblock Accessed: 2025-10-03.

\bibitem[{Mienye et~al.(2024)Mienye, Swart, and Obaido}]{info15090517}
Ibomoiye~Domor Mienye, Theo~G. Swart, and George Obaido. 2024.
\newblock \href {https://doi.org/10.3390/info15090517} {Recurrent neural networks: A comprehensive review of architectures, variants, and applications}.
\newblock \emph{Information}, 15(9).

\bibitem[{Minsky(1986)}]{minsky1986society}
Marvin Minsky. 1986.
\newblock \emph{The society of mind}.
\newblock Simon \& Schuster.

\bibitem[{Ogueji et~al.(2021)Ogueji, Zhu, and Lin}]{specialization_african_afriberta}
Kelechi Ogueji, Yuxin Zhu, and Jimmy Lin. 2021.
\newblock Small data? no problem! exploring the viability of pretrained multilingual language models for low-resourced languages.
\newblock In \emph{Proceedings of the 1st workshop on multilingual representation learning}, pages 116--126.

\bibitem[{Ouyang et~al.(2022)Ouyang, Wu, Jiang, Almeida, Wainwright, Mishkin, Zhang, Agarwal, Slama, Ray et~al.}]{ouyang2022training}
Long Ouyang, Jeff Wu, Xu~Jiang, Diogo Almeida, Carroll Wainwright, Pamela Mishkin, Chong Zhang, Sandhini Agarwal, Katarina Slama, Alex Ray, and 1 others. 2022.
\newblock Training language models to follow instructions with human feedback.
\newblock \emph{Advances in Neural Information Processing Systems}, 35:27730--27744.

\bibitem[{Park et~al.(2023)Park, O'Brien, Cai, Morris, Liang, and Bernstein}]{park2023generative}
Joon~Sung Park, Joseph~C O'Brien, Carrie~J Cai, Meredith~Ringel Morris, Percy Liang, and Michael~S Bernstein. 2023.
\newblock Generative agents: Interactive simulacra of human behavior.
\newblock \emph{Proceedings of the 36th Annual ACM Symposium on User Interface Software and Technology}.

\bibitem[{Pfeiffer et~al.(2023)Pfeiffer, Ruder, Vuli{\'c}, and Ponti}]{pfeiffer2023modular}
Jonas Pfeiffer, Sebastian Ruder, Ivan Vuli{\'c}, and Edoardo~Maria Ponti. 2023.
\newblock Modular deep learning.
\newblock \emph{arXiv preprint arXiv:2302.11529}.

\bibitem[{{Phenx AI}(2025)}]{phenxai2025routing}
{Phenx AI}. 2025.
\newblock The art of traffic control: Mastering llm routing.
\newblock \url{https://phenx.ai/}.
\newblock Accessed: 2025-05-21.

\bibitem[{{PMC}(2025)}]{pmc2025explainable}
{PMC}. 2025.
\newblock Editorial: Explainable ai in natural language processing.
\newblock \url{https://www.ncbi.nlm.nih.gov/pmc/}.
\newblock Accessed: 2025-05-21.

\bibitem[{{Prem Blog}(2025{\natexlab{a}})}]{prem2025balancing}
{Prem Blog}. 2025{\natexlab{a}}.
\newblock Balancing llm costs and performance: A guide to smart deployment.
\newblock \url{https://premai.io/blog/}.
\newblock Accessed: 2025-05-21.

\bibitem[{{Prem Blog}(2025{\natexlab{b}})}]{prem2025routing}
{Prem Blog}. 2025{\natexlab{b}}.
\newblock Llm routing: Costs optimisation without sacrificing quality.
\newblock \url{https://premai.io/blog/}.
\newblock Accessed: 2025-05-21.

\bibitem[{{Premai Blog}(2025)}]{premai_edge_2025}
{Premai Blog}. 2025.
\newblock \href {https://blog.premai.io/small-language-models-slms-for-efficient-edge-deployment/} {Small language models ({SLMs}) for efficient edge deployment}.
\newblock Blog post.
\newblock Accessed: 2025-03-04.

\bibitem[{Quirke et~al.(2025)Quirke, Neo, and Barez}]{specialization_reasoning_arithmetic_quirke}
Philip Quirke, Clement Neo, and Fazl Barez. 2025.
\newblock \href {https://doi.org/10.48550/arXiv.2402.02619} {Understanding addition and subtraction in transformers}.

\bibitem[{Ricardo(1817)}]{ricardo1817principles}
David Ricardo. 1817.
\newblock \emph{On the principles of political economy and taxation}.
\newblock John Murray, London.

\bibitem[{Riquelme et~al.(2021)Riquelme, Puigcerver, Mustafa, Neumann, Jenatton, Susano~Pinto, Keysers, and Houlsby}]{riquelme2021scaling}
Carlos Riquelme, Joan Puigcerver, Basil Mustafa, Maxim Neumann, Rodolphe Jenatton, Andr{\'e} Susano~Pinto, Daniel Keysers, and Neil Houlsby. 2021.
\newblock Scaling vision with sparse mixture of experts.
\newblock \emph{Advances in Neural Information Processing Systems}, 34:8583--8595.

\bibitem[{R{\"u}ffler et~al.(2012)R{\"u}ffler, Hermisson, and Wagner}]{rueffler2012evolution}
Claus R{\"u}ffler, Joachim Hermisson, and G{\"u}nter~P Wagner. 2012.
\newblock Evolution of functional specialization and division of labor.
\newblock \emph{Proceedings of the National Academy of Sciences}, 109(6):E326--E335.

\bibitem[{Rushby(1994)}]{rushby1994critical}
John Rushby. 1994.
\newblock Critical system properties: Survey and taxonomy.
\newblock \emph{Reliability Engineering \& System Safety}, 43(2):189--219.

\bibitem[{Saunders and DeNeefe(2024)}]{forgetting_2}
Danielle Saunders and Steve DeNeefe. 2024.
\newblock Domain adapted machine translation: What does catastrophic forgetting forget and why?
\newblock \emph{arXiv preprint arXiv:2412.17537}.

\bibitem[{Shao and Feng(2022)}]{forgetting_1}
Chenze Shao and Yang Feng. 2022.
\newblock Overcoming catastrophic forgetting beyond continual learning: Balanced training for neural machine translation.
\newblock \emph{arXiv preprint arXiv:2203.03910}.

\bibitem[{Shnitzer et~al.(2024)Shnitzer, Lin, Yang, Cheng, Begnaud, Zhang, Cheng, Chatterjee, and Jin}]{shnitzer2024universal}
Tal Shnitzer, Zichang Lin, Cheng Yang, Hao Cheng, Shelby Begnaud, Tianyi Zhang, Heng-Tze Cheng, Soham Chatterjee, and Hongyang~R Jin. 2024.
\newblock Universal model router improves llm performance across diverse benchmarks.
\newblock \emph{arXiv preprint arXiv:2502.08773}.

\bibitem[{Shu et~al.(2024)Shu, Zhao, Liu, Demeter, Du, and Zhang}]{specialization_legal_lawllm}
Dong Shu, Haoran Zhao, Xukun Liu, David Demeter, Mengnan Du, and Yongfeng Zhang. 2024.
\newblock Lawllm: Law large language model for the us legal system.
\newblock pages 4882--4889.

\bibitem[{Shumailov et~al.(2024)Shumailov, Shumailov, Zhao, Gal, Papernot, and Anderson}]{shumailov2023curse}
Ilia Shumailov, Zakhar Shumailov, Yiren Zhao, Yarin Gal, Nicolas Papernot, and Ross Anderson. 2024.
\newblock The curse of recursion: Training on generated data makes models forget.
\newblock \emph{arXiv preprint arXiv:2305.17493}.

\bibitem[{Simhi et~al.(2025)Simhi, Itzhak, Barez, Stanovsky, and Belinkov}]{simhi2025trust}
Adi Simhi, Itay Itzhak, Fazl Barez, Gabriel Stanovsky, and Yonatan Belinkov. 2025.
\newblock \href {https://arxiv.org/abs/2502.12964} {Trust me, i'm wrong: High-certainty hallucinations in llms}.
\newblock \emph{arXiv preprint arXiv:2502.12964}.

\bibitem[{Singh et~al.(2025)Singh, Magazine, Pandya, and Nambi}]{singh2025artist}
Joykirat Singh, Raghav Magazine, Yash Pandya, and Akshay Nambi. 2025.
\newblock \href {https://arxiv.org/abs/2505.01441} {Agentic reasoning and tool integration for llms via reinforcement learning}.
\newblock \emph{arXiv preprint arXiv:2505.01441}.

\bibitem[{Smith(1776)}]{smith1776wealth}
Adam Smith. 1776.
\newblock \emph{An inquiry into the nature and causes of the wealth of nations}.
\newblock W. Strahan and T. Cadell, London.

\bibitem[{{SmythOS}(2025)}]{smythos2025explainable}
{SmythOS}. 2025.
\newblock Explainable ai in natural language processing: Enhancing transparency and trust in language models.
\newblock \url{https://smythos.com/}.
\newblock Accessed: 2025-05-21.

\bibitem[{Somerstep et~al.(2025)Somerstep, Polo, de~Oliveira, Mangal, Silva, Bhardwaj, Yurochkin, and Maity}]{carrot_router}
Seamus Somerstep, Felipe~Maia Polo, Allysson Flavio~Melo de~Oliveira, Pratyush Mangal, Mirian Silva, Onkar Bhardwaj, Mikhail Yurochkin, and Subha Maity. 2025.
\newblock Carrot: A cost aware rate optimal router.
\newblock In \emph{ICLR 2025 Workshop on Foundation Models in the Wild}.
\newblock \url{https://huggingface.co/CARROT-LLM-Routing}.

\bibitem[{Stickland et~al.(2024)Stickland, Lyzhov, Pfau, Mahdi, and Bowman}]{steering_side_effects}
Asa~Cooper Stickland, Alexander Lyzhov, Jacob Pfau, Salsabila Mahdi, and Samuel~R Bowman. 2024.
\newblock Steering without side effects: Improving post-deployment control of language models.
\newblock \emph{arXiv preprint arXiv:2406.15518}.

\bibitem[{Storey(1996)}]{storey1996safety}
Neil Storey. 1996.
\newblock \emph{Safety-Critical Computer Systems}.
\newblock Addison-Wesley, Harlow, England.

\bibitem[{Team(2025{\natexlab{a}})}]{llm_evaluation_metrics}
Aisera~Research Team. 2025{\natexlab{a}}.
\newblock \href {https://aisera.com/blog/llm-evaluation/} {Llm evaluation: Key metrics, best practices and frameworks}.
\newblock Technical report, Aisera.

\bibitem[{Team(2025{\natexlab{b}})}]{ai_fairness_metrics}
Forbes Councils~Technology Team. 2025{\natexlab{b}}.
\newblock \href {https://councils.forbes.com/blog/ai-and-fairness-metrics} {Ai \& fairness metrics: Understanding \& eliminating bias}.
\newblock \emph{Forbes Councils}.

\bibitem[{Team(2025{\natexlab{c}})}]{llm_bias_testing}
Test.io~Research Team. 2025{\natexlab{c}}.
\newblock \href {https://academy.test.io/en/articles/9227500-llm-bias-understanding-mitigating-and-testing-the-bias-in-large-language-models} {Llm bias: Understanding, mitigating and testing the bias in large language models}.
\newblock Technical report, Test.io.

\bibitem[{Thakur et~al.(2024)Thakur, Choudhary, Ramayapally, Vaidyanathan, and Hupkes}]{thakur2024judging}
Aman~Singh Thakur, Kartik Choudhary, Venkat~Srinik Ramayapally, Sankaran Vaidyanathan, and Dieuwke Hupkes. 2024.
\newblock Judging the judges: Evaluating alignment and vulnerabilities in llms-as-judges.
\newblock \emph{arXiv preprint arXiv:2406.12624}.

\bibitem[{Tiwari(2024)}]{tiwari_edge_2024}
Pankaj Tiwari. 2024.
\newblock \href {https://medium.com/accredian/edge-ai-deploying-large-language-models-for-smarter-devices-cdee25023673} {Edge {AI}: Deploying large language models for smarter devices}.
\newblock Medium: Accredian.
\newblock Accessed: 2024-09-13.

\bibitem[{{Towards Data Science}(2025)}]{towardsdatascience2025routing}
{Towards Data Science}. 2025.
\newblock Llm routing—intuitively and exhaustively explained.
\newblock \url{https://towardsdatascience.com/}.
\newblock Accessed: 2025-05-21.

\bibitem[{Varoquaux et~al.(2025)Varoquaux, Luccioni, and Whittaker}]{varoquaux2025hype}
Gaël Varoquaux, Alexandra~Sasha Luccioni, and Meredith Whittaker. 2025.
\newblock \href {https://arxiv.org/abs/2409.14160} {Hype, sustainability, and the price of the bigger-is-better paradigm in ai}.
\newblock \emph{arXiv preprint arXiv:2409.14160}.

\bibitem[{Walzer(2008)}]{walzer2008spheres}
Michael Walzer. 2008.
\newblock \emph{Spheres of justice: A defense of pluralism and equality}.
\newblock Basic books.

\bibitem[{Wu(2011)}]{master_switch}
Tim Wu. 2011.
\newblock \emph{The master switch: The rise and fall of information empires}.
\newblock Vintage.

\bibitem[{Xie et~al.(2024)Xie, Chen, Chen, Peng, Hu, Lin, Peng, Huang, Zhang, Keloth et~al.}]{specialization_medical_mellama}
Qianqian Xie, Qingyu Chen, Aokun Chen, Cheng Peng, Yan Hu, Fongci Lin, Xueqing Peng, Jimin Huang, Jeffrey Zhang, Vipina Keloth, and 1 others. 2024.
\newblock Me-llama: Foundation large language models for medical applications.
\newblock \emph{Research square}, pages rs--3.

\bibitem[{Xu et~al.(2020)Xu, Zhong, Yepes, and Lau}]{forgetting_3}
Ying Xu, Xu~Zhong, Antonio Jose~Jimeno Yepes, and Jey~Han Lau. 2020.
\newblock Forget me not: Reducing catastrophic forgetting for domain adaptation in reading comprehension.
\newblock In \emph{2020 International joint conference on neural networks (IJCNN)}, pages 1--8. IEEE.

\bibitem[{Yang et~al.(2024)Yang, Shen, Guo, Wang, Cao, Zhang, and Tao}]{Survery_ModelMerging_2024}
Enneng Yang, Li~Shen, Guibing Guo, Xingwei Wang, Xiaochun Cao, Jie Zhang, and Dacheng Tao. 2024.
\newblock Model merging in llms, mllms, and beyond: Methods, theories, applications and opportunities.
\newblock \emph{arXiv preprint arXiv:2408.07666}.

\bibitem[{Yang et~al.(2025)Yang, Zhang, Song, Hong, Xu, Zhao, Zhang, Cui, and Yang}]{yang2025diffusionmodelscomprehensivesurvey}
Ling Yang, Zhilong Zhang, Yang Song, Shenda Hong, Runsheng Xu, Yue Zhao, Wentao Zhang, Bin Cui, and Ming-Hsuan Yang. 2025.
\newblock \href {https://arxiv.org/abs/2209.00796} {Diffusion models: A comprehensive survey of methods and applications}.
\newblock \emph{Preprint}, arXiv:2209.00796.

\bibitem[{Ye et~al.(2025)Ye, Tang, Ge, Du, Yin, Chen, and Shao}]{ye2025masgpt}
Rui Ye, Shuo Tang, Rui Ge, Yaxin Du, Zhenfei Yin, Siheng Chen, and Jing Shao. 2025.
\newblock \href {https://github.com/rui-ye/MAS-GPT} {Mas-gpt: Training llms to build llm-based multi-agent systems}.
\newblock In \emph{International Conference on Machine Learning (ICML)}.
\newblock Poster.

\bibitem[{Zaharia et~al.(2024)Zaharia, Khattab, Chen, Davis, Miller, Potts, Zou, Carbin, Frankle, Rao, and Ghodsi}]{compound-ai-blog}
Matei Zaharia, Omar Khattab, Lingjiao Chen, Jared~Quincy Davis, Heather Miller, Chris Potts, James Zou, Michael Carbin, Jonathan Frankle, Naveen Rao, and Ali Ghodsi. 2024.
\newblock The shift from models to compound ai systems.
\newblock \url{https://bair.berkeley.edu/blog/2024/02/18/compound-ai-systems/}.

\bibitem[{Zhang et~al.(2023)Zhang, Li, Cui, Cai, Liu, Fu, Huang, Zhao, Zhang, Chen, Wang, Luu, Bi, Shi, and Shi}]{zhang2023sirens}
Yue Zhang, Yafu Li, Leyang Cui, Deng Cai, Lemao Liu, Tingchen Fu, Xinting Huang, Enbo Zhao, Yu~Zhang, Yulong Chen, Longyue Wang, Anh~Tuan Luu, Wei Bi, Freda Shi, and Shuming Shi. 2023.
\newblock Siren's song in the ai ocean: A survey on hallucination in large language models.
\newblock \emph{arXiv preprint arXiv:2309.01219}.

\bibitem[{Zhao et~al.(2024)Zhao, Wang, Zhang, Han, Deveci, and Parmar}]{zhao_review_2024}
Xia Zhao, Limin Wang, Yufei Zhang, Xuming Han, Muhammet Deveci, and Milan Parmar. 2024.
\newblock \href {https://doi.org/10.1007/s10462-024-10721-6} {A review of convolutional neural networks in computer vision}.
\newblock \emph{Artificial Intelligence Review}, 57(4):99.

\bibitem[{Zhong et~al.(2024)Zhong, Liang, Wang, Wang, Huang, and Li}]{zhong2024adapmoe}
Shuzhang Zhong, Ling Liang, Yuan Wang, Runsheng Wang, Ru~Huang, and Meng Li. 2024.
\newblock \href {https://arxiv.org/abs/2408.10284} {Adapmoe: Adaptive sensitivity-based expert gating and management for efficient moe inference}.
\newblock \emph{arXiv preprint arXiv:2408.10284}.

\bibitem[{Zhou et~al.(2020)Zhou, Cui, Hu, Zhang, Yang, Liu, Wang, Li, and Sun}]{ZHOU202057}
Jie Zhou, Ganqu Cui, Shengding Hu, Zhengyan Zhang, Cheng Yang, Zhiyuan Liu, Lifeng Wang, Changcheng Li, and Maosong Sun. 2020.
\newblock \href {https://doi.org/10.1016/j.aiopen.2021.01.001} {Graph neural networks: A review of methods and applications}.
\newblock \emph{AI Open}, 1:57--81.

\end{thebibliography}

\appendix

\section{Appendix A: Theoretical Foundations for Expert Orchestration}
\label{app:TheoreticalFoundations}

This appendix provides detailed theoretical support for the Expert Orchestration framework from multiple disciplines.

\para{A.1 Economic Theories of Distributed Knowledge and Market Structure}

Friedrich Hayek's seminal work on distributed knowledge \citep{hayek1945use} provides a powerful economic framework supporting EO. Hayek argued that knowledge in society exists as ``dispersed bits of incomplete and frequently contradictory knowledge which all the separate individuals possess," never in ``concentrated or integrated form" in any single mind. This impossibility of centralizing all knowledge leads to the superiority of market mechanisms over central planning: markets function as information processors that coordinate distributed expertise through price signals.

Adam Smith's theory of the division of labor \citep{smith1776wealth}  illustrates how breaking complex tasks into specialized functions dramatically increases productivity. Smith further observed that ``the division of labor is limited by the extent of the market", meaning specialization increases as markets grow. This principle applies directly to models: as the demand for capabilities expands, we should expect greater specialization of models rather than continued focus on general-purpose systems.

David Ricardo's theory of comparative advantage \citep{ricardo1817principles} extends this insight, showing that even when one agent is superior at all tasks, the total output is maximized if agents specialize according to their relative strengths.

\para{A.2 Organizational Theory: Collective Decision-Making and Diversity}

Condorcet's Jury Theorem \citep{condorcet1785essay} provides mathematical proof that groups of independent decision-makers with better-than-random accuracy consistently outperform individuals, with reliability approaching certainty as group size grows. This applies directly to EO, where specialized judge models serve as an ``expert jury" providing more reliable assessment than any single generalist model.

Lu Hong and Scott Page's diversity theorem \citep{hong2004groups}  extends this insight, proving that ``groups of diverse problem solvers can outperform groups of high-ability problem solvers." EO leverages this principle by maintaining diverse specialized models, each bringing distinct problem-solving approaches to user queries.

\para{A.3 Cognitive Science: Modularity and Distributed Intelligence}

Cognitive science provides compelling evidence that intelligence naturally emerges from specialized, interacting components rather than monolithic processors. Jerry Fodor's ``Modularity of Mind" theory \citep{fodor1983modularity} demonstrates that human cognition comprises domain-specific modules specialized for particular functions like language or vision, each operating with some independence from others. This modularity enables both efficiency and robustness—when one module fails, others continue functioning.

Building on this foundation, Marvin Minsky's ``Society of Mind" theory \citep{minsky1986society} offers a direct parallel to EO. Minsky proposed that intelligence emerges from ``the interaction of many small, simple parts" without requiring a complex central controller: ``a model of the human mind more like a democracy than a supercomputer." Recent AI research has validated this approach: \citet{park2023generative} demonstrated that over a hundred specialized LLM agents working together can outperform any single model on complex tasks by collaborating and sharing information.

\para{A.4 Biological and Evolutionary Frameworks}

The evolution of multicellular life provides a compelling analogy for EO. Single-celled organisms function as generalists, handling all life processes internally. The transition to multicellularity involved cells specializing into different types (muscle, nerve, blood, etc.), dramatically increasing the organism's capabilities. As Rüffler et al. note, "division of labor among functionally specialized modules occurs at all levels of biological organization" and represents a major evolutionary trend because specialization enables higher performance \citet{rueffler2012evolution}.

\para{A.5 Routing as a Form of Democratic Algorithmic Institution}

EO reflects key democratic values: participation, accountability, and distributed influence. Robert Dahl emphasizes that democracy depends on broad inclusion and equal ability to shape outcomes \citep{dahl2008democracy}, while Jürgen Habermas underscores the role of open, reasoned dialogue in legitimizing decisions \citep{habermas2015between}. John Dewey sees democracy as collective problem-solving rooted in everyday association \citep{dewey2012public}.

EO echoes these ideals by lowering barriers for niche model creators, enabling a wider range of contributors to offer specialized capabilities. Through open evaluation and fair task routing, it promotes meaningful participation and healthy competition. Like the U.S. system of checks and balances, this distribution of influence helps prevent dominance by any single actor, fosters fairness, and supports systemic stability (\citep{madison1788federalist}. This pluralistic structure enables excellence across diverse domains and interests—an ideal at the heart of Walzer's argument for justice through distinct but coexisting spheres of merit \citep{walzer2008spheres}.

\para{A.6 Alignment and Safety Approaches}

The safety via debate framework \citep{irving2018ai} proposes training agents to engage in adversarial debates about questions, with a human or judge model determining which agent provides the most convincing answer. This approach uses multiple systems with potentially opposed viewpoints to surface flaws in each other's reasoning, improving the trustworthiness of answers. EO naturally incorporates this debate-like structure through its judge models.

\citet{christiano2018supervising}'s Iterated Distillation and Amplification (IDA) alignment framework parallels EO principles. IDA starts with humans or simple models breaking complex tasks into smaller sub-questions, answering those questions, and then aggregating the answers. This decomposition approach is then distilled into a more efficient model, which is iteratively amplified through additional decomposition.

\para{A.7 Synthesis}

Across economics, cognition, biology, and organizational theory, specialized coordinated systems consistently outperform monolithic designs for complex tasks. From Hayek's distributed knowledge to Minsky's society of mind to multicellular evolution, the pattern is clear: complex capabilities emerge through orchestrated interaction of specialized components, not through scaling generalist systems. EO applies these proven principles to AI, creating systems that are more capable, transparent, and democratically governable than monolithic alternatives.

\end{document}